\documentclass[aps,pre,twocolumn,showpacs,superscriptaddress,groupedaddress,nofootinbib]{revtex4}  

\usepackage{graphicx}
\usepackage{epstopdf}
\usepackage{bm}
\usepackage{amsmath,amsfonts,amssymb,amstext}
\usepackage[utf8]{inputenc}
\usepackage{color}

\newcommand{\be}{\begin{equation}}
\newcommand{\ee}{\end{equation}}
\newcommand{\bea}{\begin{eqnarray} \nonumber }
\newcommand{\eea}{\end{eqnarray}}
\newcommand{\bi}{\begin{itemize}}
\newcommand{\ei}{\end{itemize}}

\newcommand{\Me}{M\!e}
\newcommand{\vectr}{\bm{r}}
\newcommand{\vpara}{\bm{v}_{/\!/}} 

\newcommand{\vectfv}{\bm{F}_v}
\newcommand{\vectF}{\bm{F}}

\newcommand{\affcergy}{\address{Laboratoire de Physique Th\'{e}orique et Mod\'{e}lisation, Universit\'{e} de Cergy Pontoise, CNRS-UMR~8089, 2 avenue A. Chauvin, 95302 Cergy-Pontoise, France}}
\newcommand{\afflangevin}{\address{Institut Langevin, ESPCI ParisTech, CNRS-UMR~7587, 1 rue Jussieu, 75005 Paris Cedex 05, France}}
\newcommand{\affmsc}{\address{Mati\`{e}re et Syst\`{e}mes Complexes, Universit\'{e} Paris Diderot, CNRS-UMR~7057, 10 rue A. Domon et L. Duquet, 75013 Paris, France}}
\newcommand{\affLPPMH}{\address{Laboratoire de Physique et M\'{e}canique des Milieux H\'{e}t\'{e}rog\`{e}nes, ESPCI ParisTech, CNRS-UMR~7636, 10 rue Vauquelin, 75005 Paris, France}}

\begin{document}


\title{Interaction of two walkers: Wave-mediated energy and force}

\author{Christian Borghesi}
\thanks {christian.borghesi@u-cergy.fr}
\affcergy

\author{Julien Moukhtar}
\affmsc

\author{Matthieu Labousse}
\afflangevin

\author{Antonin Eddi}
\affLPPMH

\author{Emmanuel Fort}
\afflangevin

\author{Yves Couder}
\affmsc



\date{\today}

\begin{abstract}
A bouncing droplet, self-propelled by its interaction with the waves it generates, forms a classical wave-particle association called a ``walker." Previous works have demonstrated that the dynamics of a single walker is driven by its global surface wave field that retains information on its past trajectory. Here, we investigate the energy stored in this wave field for two coupled walkers and how it conveys an interaction between them. For this purpose, we characterize experimentally the ``promenade modes'' where two walkers are bound, and propagate together. Their possible binding distances take discrete values, and the velocity of the pair depends on their mutual binding. The mean parallel motion can be either rectilinear or oscillating. The experimental results are recovered analytically with a simple theoretical framework. A relation between the kinetic energy of the droplets and the total energy of the standing waves is established. 

\end{abstract}


\maketitle

\section{Introduction}
The specific dynamical properties of a \textit{walker} result from a wave-mediated self-organization. In this particle-wave association, the drop generates a standing surface field and this wave field pilots the particle motion. Several previous works have revealed that this system exhibits dual properties~\cite{JFM06,diffraction-interference,exotic-orbites,tunnel,path-memory-pnas,JFM11,zeeman,cavite,Dan-rotating,SO-eingenstates}. They were \textit{hitherto} investigated through the observed dynamics of the particle. Here, we wish to rationalize the same system from the viewpoint of the guiding wave, its stored energy, and the forces it generates. We use the existence of bound states to address this problem.

A single walker is obtained when a droplet, placed on a vibrated bath, bounces at a sub-harmonic frequency and is thus a source of Faraday standing waves. It then becomes self-propelled by its interaction with these waves~\cite{JFM06}. The resulting wave-particle association is a dissipative structure sustained by the external imposed vibration. Two walkers coexisting on the same bath are known to have a long-range interaction when their field overlaps. It was shown that the collision of two counter-propagating identical walkers may lead to the formation of orbiting bound states having a discrete set of possible diameters~\cite{JFM06,exotic-orbites,pilot-wave-hydro}. As mentioned in earlier articles~\cite{JFM06,pilot-wave-hydro}, other modes of self-organization of two drops are observed in which a bound pair of walkers, moves on parallel trajectories. In this type of motion that we called the ``promenade modes" the two drops can move either rectilinearly in parallel or on low-frequency symmetrically oscillating trajectories. In the present article, because of their geometrical simplicity, we use these promenade modes for a preliminary investigation of the energy of walkers. Figure~\ref{fh-exp} shows the difference between the wave field of a single walker (a) and that of pair of walkers bound in several promenade modes (b-d). Here, we seek to relate directly the effect of the wave interference observed in panels (b)-(d) with the dynamics of the bound mode.

\section{Description of a coupled system: the two-dimensional motion in the promenade mode}
\subsection{Experimental set-up \label{sexperiment}}

\begin{figure}
\centering
\includegraphics[width=1 \columnwidth, clip=true]{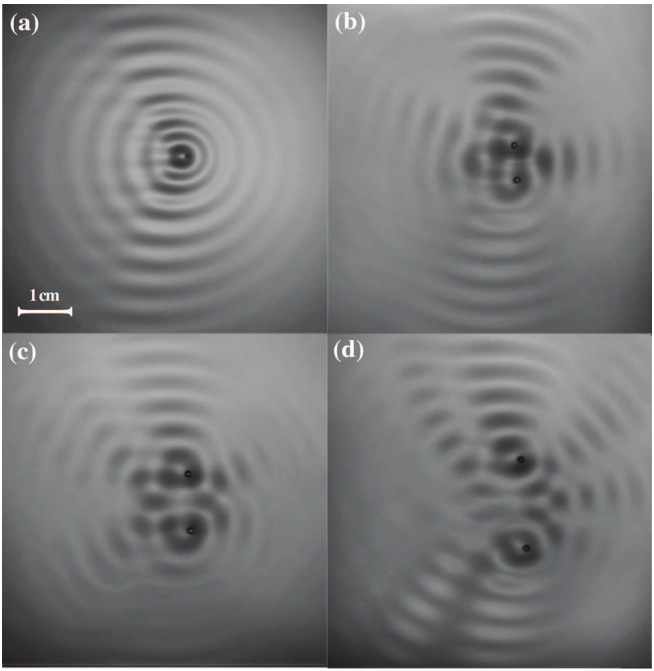}
\caption{\small Photographs of the wave fields. (a) A single walker of velocity $V_0 = 9\:\mathrm{mm/s}$ at a memory $\Me = 100$. (b-c) The wave fields of two droplets bound into promenade modes $n=1$ (b), $n=2$ (c) and $n=3$ (d) at a memory $\Me = 50$. The ratio of the velocities of the bound pairs to the free velocity of the droplets are, respectively, 0.55 ($n=1$), 0.75 ($n=2$) and 0.85 ($n=3$).}
\label{fh-exp}
\end{figure}

The experimental set up is identical to that described in Ref. \cite{JFM11}. A tank filled with silicon oil (of viscosity $\mu=20$~cP) is oscillated vertically at a frequency $f_0=80$~Hz with an acceleration $\gamma=\gamma_m \sin(2\pi f_0 t)$. The amplitude $\gamma_m$ can be continuously tuned from a value of the order of the acceleration of gravity $g$ up to the Faraday instability threshold observed at $\gamma_m^F=4.3\,g$. The \textit{walking regime} appears at a threshold $\gamma_m^w\approx 3.2\,g$ (with $\gamma_m^w<\gamma_m^F$), when the drop of mass $m_d$ becomes a source of damped Faraday waves (\cite{JFM11,molacek2}) of wavelength $\lambda_F=4.75$~mm. While it increases immediately above the threshold $\gamma_m^w$, in most of the walking regime the drop velocity saturates at a constant value $V_0$ that depends on the drop size. We limited our investigation to situations where the two walkers were identical in which the droplets have the same size and the same free velocity $V_0$. This is a requirement for having a pair walking in parallel. We thus used several pairs of identical droplets having diameters in the range $0.7\leq D \leq 0.75\,\mathrm{mm}$ ({\it i.e.}, masses $0.17\leq m_d\leq 0.22\,\mathrm{mg}$). Their free velocities $V_0$, were in the range $6.5\leq V_0 \leq 10.5\,\mathrm{mm/s}$. We controlled an initial state by sending two walkers to a collision region with initial velocities at a small angle from each other. In such cases, the two drops bound to each other so that after the collision their trajectories were either parallel or symmetrically oscillating at a low frequency. In most situations the bound pair oscillates. The typical life-time of a bound pair is usually short because the promenade mode is fragile: the bound pair of walkers is usually disrupted by unavoidable collisions with the cell walls. In the case where the pair has a parallel motion it is not clear whether it would not start oscillating after a finite time. It is possible to detach and bind the drops repeatedly and, by varying the parameters of the initial collision angle, to obtain a large number of different binding modes with the same pair of droplets. Note, however, that two situations can exist. The bouncing droplets being sub-harmonic, two drops can bounce either synchronously or with opposite phases. Going from one situation to the other requires disturbing the bouncing of one of the drops. 

\begin{figure}
\centering
\includegraphics[width=7cm,height=10cm,clip=true]{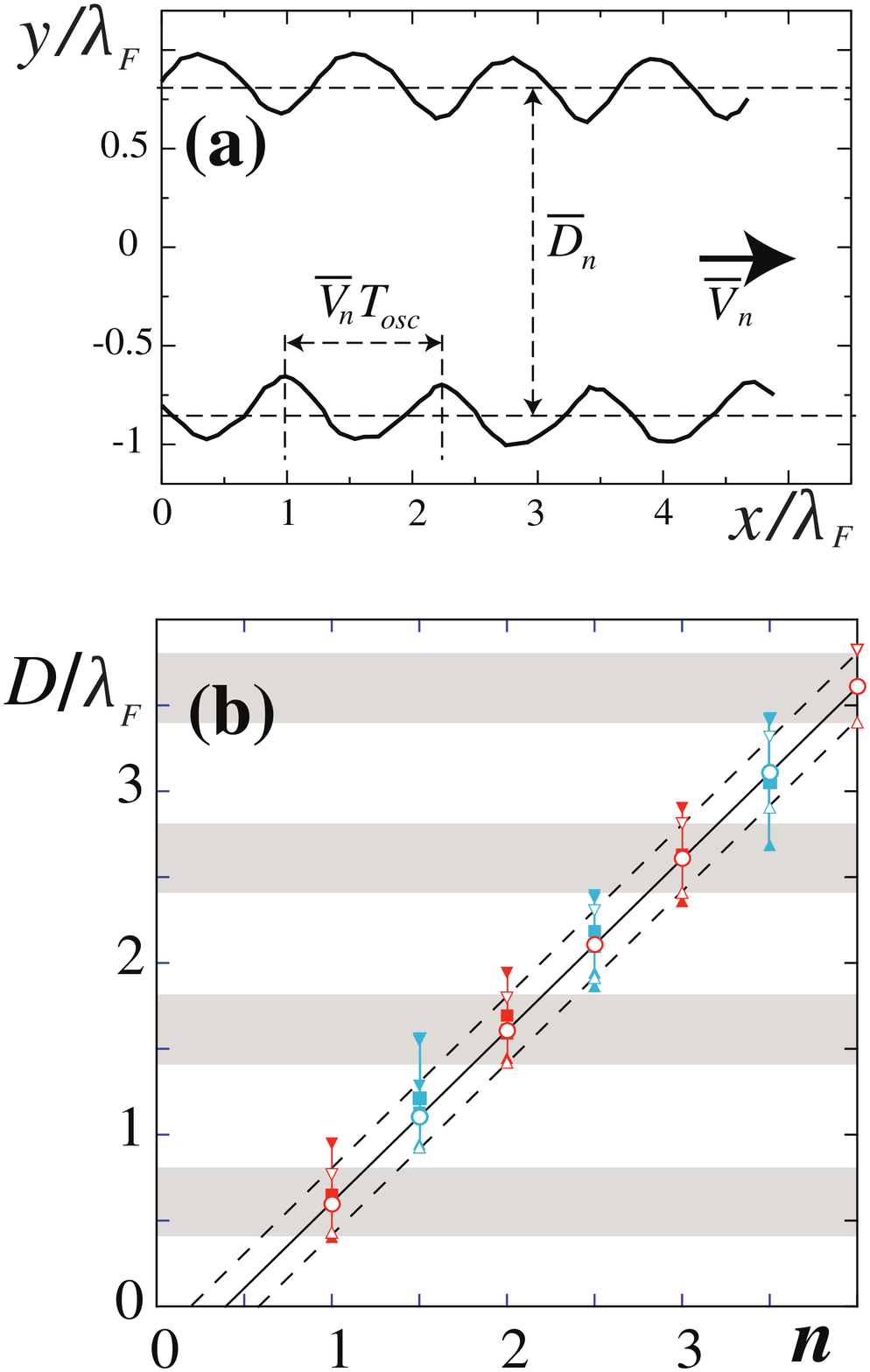}
\caption{\small (Color online) Characterization of the promenade modes. (a) Trajectories of the two droplets for the mode $n=2$. The average distance between the two drops is $\overline{D}_2\approx 1.6\,\lambda_F$, the steady velocity along the $x$ axis is $\overline{V}_2\approx 7.5\:\mathrm{mm/s}$, and the oscillation period $T_{osc}\approx 30\: T_F$. (b) The discrete possible distances of binding $\overline{D}_n/\lambda_F$. The experimental mean binding distances of drops bouncing in phase are shown as red squares, of opposite phase as blue squares. The graph also shows (triangles) the maximal and minimal distances observed during the oscillations. The open circles are the theoretical predicted binding distances. The shaded areas show the allowed distances between droplets when they bounce in phase.}
\label{fexp-1}
\end{figure}

\begin{figure}
\centering
\includegraphics[width=7cm,height=10cm,clip=true]{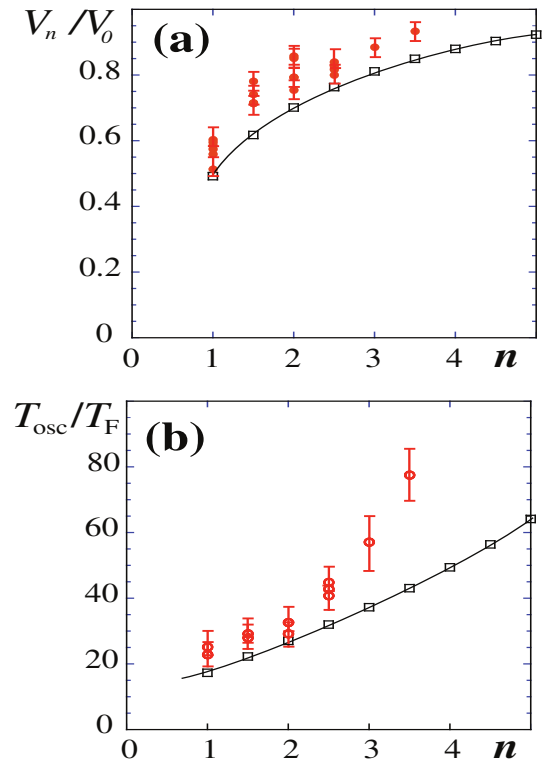}
\caption{\small (Color online) The velocities and oscillations of the promenade modes. (a) The experimentally measured ratios $\overline{V}_n/V_0$ of the steady mean velocity of the bound pair to the velocity of the same droplets when unbound (red open circles). (b) Measured ratio $T_{osc}/T_F$ of the oscillation period, to the bouncing period, for different modes (red open circles). In both (a) and (b) the black open squares are the corresponding analytical results.}
\label{fexp-2}
\end{figure}

From the experimental recordings of the droplet motion~\cite{JFM06,JFM11}, we first determine the trajectories. An example is shown in Fig.~\ref{fexp-1}(a). We measure the mean binding distances $\overline{D}_n$. As shown in Fig.~\ref{fexp-1}(b) they form a discrete set of values directly related to the Faraday wavelength. An empirical fit of these mean distances gives
\begin{equation}
\overline{D}_n = (n-\epsilon_0)\lambda_F.
\end{equation}
For drops bouncing in phase, $n$ are the successive integers $n=1,\, 2,\, 3,\dots$ with opposite phases the successive values of $n$ are: $3/2,\, 5/2,\dots$. The apparent offset $\epsilon_0 = 0.32 \pm 0.02$ is practically the same for all the modes~\cite{exotic-orbites}. These distances $\overline{D}_n$ are close but slightly smaller than the diameters observed for the orbiting bound states of the same drops (for which \cite{JFM06} $\epsilon_0 = 0.2 \pm 0.05$). However we note that no promenade mode is observed at the short distance corresponding to the tightest orbit $n=1/2$. The extremal distances $D_n^{\mathrm{min}}$ and $D_n^{\mathrm{max}}$ separating the drops in the oscillating promenade mode are also shown in Fig.~\ref{fexp-1}(a) and (b). When two drops are bound in a promenade mode, their mean velocity $\overline{V}_n$ in the direction of propagation is reduced compared to the free velocity $V_0$ of the same drops. The evolution of the ratio $\overline{V}_n/V_0$ with $n$ [Fig.~\ref{fexp-2}(a)] shows that the tighter the bound, the slower the velocity. A similar trend is observed for the orbiting modes of two droplets. Finally we also measured the transverse oscillation period [Fig.~\ref{fexp-2}(b)]; it is remarkably large as compared to the bouncing period $T_F$. We note that the tighter the binding of the drops, the smaller the period of oscillation. Finally we remark that this periodicity is close to that observed in oscillating orbits~\cite{exotic-orbites}. 

Finally an additional remark involves the extent of the wave field. It has been shown in various experimental situations \cite{path-memory-pnas,JFM11} that an important parameter controlling the dynamics of walkers is what was called the ``wave-mediated path memory'' of this system. Because of the proximity of the Faraday instability threshold the waves generated by the sub-harmonic bouncing of the droplets are partly sustained by the effect of parametric forcing. As a result the typical time $\tau$ of their decay increases exponentially when $\gamma_m$ becomes close to $\gamma_m^F$. A non-dimensional memory parameter $\Me = \tau/T_F$ can be defined and its value estimated to be $\Me=\gamma_m/(\gamma_m^F - \gamma)$. It determines the number of past sources that contributes to the structure of the global wave field. In the present set of experiments $\gamma_m$ was varying in the range $3.85\leq \gamma_m \leq 4.2\,g$ which corresponds to values of $\Me$ ranging from 10 to 50. In this range there is no strong influence of the value of $\Me$ on the main characteristics of the trajectories. As the drops move in a straight line, the complex structure of the wave field due to the memory is left behind and does not play an important role. This is an essential difference with the experiments in which the walker is confined~\cite{Dan-rotating,SO-eingenstates}.

\subsection{Theoretical framework\label{stheoretical}}
The general theoretical framework used throughout this paper comes from Refs. \cite{path-memory-pnas,JFM11} where drops are guided by stationary waves on the bath surface. This very simple model of a single drop is mainly based on three assumptions: (1) circular standing waves are generated by the previous bounces, (2) the drop receives an incremental `kick' due to the slope of the surface bath when it collides with the fluid interface, and (3) the vertical motion of a drop is considered as independent from its horizontal motion. For the sake of simplicity, the incremental velocity `kick' is supposed to be instantaneous, when the drop hits the surface. We neglect the drop spatial extent and assimilated it to a point.

In order to write the evolution of bouncing positions $\vectr_i$ at times $t_i$ in a simple manner, we describe the discrete dynamics as generating effective forces. The details of the derivation of this model are given in the Appendix~\ref{sappendix-A}. Here we restrict ourselves to a discussion of the main results. The validity of this type of approach will be tested here in the modeling of the promenade modes. More generally it should be useful for the investigation of interacting walkers phenomena. Let $\bm{V}_{i}=(\vectr_i-\vectr_{i-1})/T_F$ be the velocity vector between two bounces of a droplet from the time labeled by $i-1$ to $i$. It is possible to write the dynamics (see the Appendix~\ref{sappendix-A}) as:
\be \label{edyn} 
\dfrac{m_d}{T_F}\left(\bm{V}_{i+1} - \bm{V}_{i}\right)= \vectfv + \vectF_W,
\ee
where $\vectfv=-(m_d/T_F)\cdot C_v \bm{V}_{i}$ is a friction-like force, and $\vectF_W=-(m_d/T_F)\cdot C_F \bm{S}_{i}$ results from the `kick' given by the slope of the wave ($\bm{S}_{i}$) to the droplet when it collides with the bath. The coefficients $C_v > 0$ and $C_F > 0$ depend on the mass of the droplet and on the interaction between the droplet and the liquid bath, and can easily be evaluated owing to~\cite{JFM11}. In our model, $C_v\approx 0.21$ and $C_F\approx 0.41\: \lambda_F/T_F$.

Each bounce, at position $\bm{r}_p$ at time $t_p$, generates a stationary circular wave. The relative surface height $h(\vectr,t_i)$, at $i$-th impact time $t_i$ and at any position $\vectr$, results from the superposition of stationary circular waves emitted from the previous bounces of one drop:
\be \displaystyle
 h(\vectr,t_i) = A \sum\limits_{p=i-N}^{i-1} e^{-(t_i-t_p)/\tau}e^{-\Vert \vectr-\vectr_p\Vert /\delta}\, \mathrm{J}_0\left[2\pi\frac{\Vert \vectr-\vectr_p\Vert}{\lambda_F}\right],
\label{eh} 
\ee
where $A$ denotes the wave amplitude at each impact, $\bm{r}_p$ stands for the positions of the previous impacts at time $t_p=t_i-(i-p)T_F$ (with $p<i$), and $\mathrm{J}_0$ indicates the Bessel function of the first kind of order $0$. $\delta$ is the typical damping distance which accounts for the viscosity of the liquid bath. ($\delta=2.5\,\lambda_F$ for numerical simulations and analytical calculations throughout this paper.) It must be recalled that at each collision a circular wave is emitted (as shown in Ref. \cite{JFM11}). If it was undamped its radial decrease would simply result from energy conservation. This wave is actually additionally spatially damped by viscosity as it propagates radially away from the point of impact. This extra damping is intrinsic and determined by the fluid viscosity (characterized by the length scale $\delta$). In the presence of vertical oscillations a packet of standing waves is generated by the Faraday effect. Its initial amplitude at a given point depends on the amplitude of the traveling wave at that point. A standing-wave pattern is thus generated with a spatially variable amplitude. Without viscosity it would be a $\mathrm{J}_0$ Bessel function, here it decreases faster radially. The memory parameter $\Me$ determines the number of past sources that have to be taken into account in the summation [Eq.~\ref{eh}]. The memory effects have been shown to be of crucial importance in the situations in which the drops are confined so that they revisit regions they have already disturbed. The memory parameter is much less critical in the situations of rectilinear motion since the ancient waves are mostly left behind. The promenade modes are observed experimentally both at low and high memory with very similar characteristics. For this reason, in the following we investigate the situations where the standing wave is only generated by the last bounce of each drop (what we call {\it low memory limit}) in the analytical part ; and low memory $\Me=4$ in the numerical simulations.

In situations of low memory, only the last bounce is taken into account as it mainly determines the wave structure, \textit{i.e.}, $N=1$ in Eq.~(\ref{eh}). The velocity of the free motion is kept as before (when $\Me=4$) by tuning $A$ in Eq.~(\ref{eh}) at a value $A_{\mathrm{eq}}$. Hence the relative surface height when a bounce collides with the surface bath is given by 
\begin{equation}
h(\vectr,t_i) = A_{\mathrm{eq}}e^{-\Vert \vectr-\vectr_{i-1}\Vert/\delta}\, \mathrm{J}_0\left[2\pi\frac{\Vert\vectr-\vectr_{i-1}\Vert}{\lambda_F}\right] + O(M^2),
\label{equationbassememoire}
\end{equation}
where $\bm{r}_{i-1}$ indicates the position of the last impact of the considered drop. 

\subsection{Validation of the model\label{sdynamics}}
\begin{figure}
\centering
\includegraphics[width=0.6 \columnwidth, clip=true]{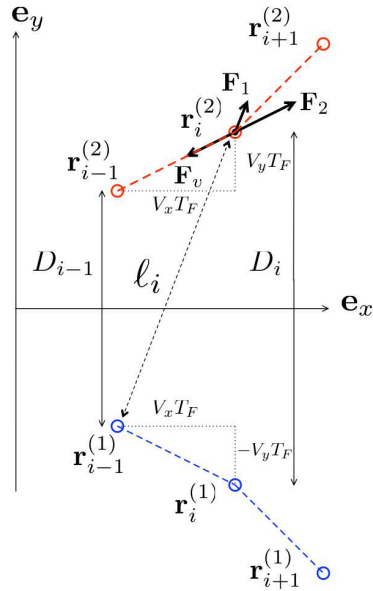}
\caption{\small (Color online) Sketch of two consecutive periods of the motion showing the forces exerted on one drop and the notations used in the computation.}
\label{fnotation}
\end{figure}
\subsubsection{Equilibrium distances}
Let us now consider the case of two identical droplets at a position $\bm{r}^{(1)}$ and $\bm{r}^{(2)}$, in promenade motion (see Fig.~\ref{fnotation}). The motion of the two droplets is assumed to be symmetrical with respect to the $x$ axis. Because of the linear superposition of the surface waves, the force $\bm{F}_W$ can be decomposed into two distinct contributions $\bm{F}_W=\bm{F}_1+\bm{F}_2$. If we consider one drop at a time $t_i$, (say, the drop 2), it reacts to the wave generated by its companion (drop 1) at its last impact, which induces an interaction force $\bm{F}_1$. It also reacts to its own standing wave generated at the single previous impact [Eq. (\ref{equationbassememoire})], which gives the self-induced driving force $\bm{F}_2$

An equilibrium distance, $D_{0,n}$, corresponds to a stable equilibrium point with regard to the force $\bm{F}_1$. Since this force is oppositely proportional to the slope where the droplet collides with the bath surface, $D_{0,n}$ is a local minimum of the relative surface height $h$, generated by one droplet at the impact position of the other one. In other words, according to Eq.~(\ref{equationbassememoire}) an equilibrium distance is a local minimum of the function $f(D)=A_{\mathrm{eq}}\, e^{-D/\delta}\,\mathrm{J}_0(2\pi\,D/\lambda_F)$. Hence,
\be
D_{0,n}/\lambda_F \approx 0.6,\; 1.6,\; 2.6,\dots ~ (n=1,\;2,\;3,\dots).
\ee
Considering two drops bouncing with opposite phases is straightforward. In this case, we consider that the relative surface height generated by one droplet at the position of the second one is in opposite sign compared to the case where droplets bounce in phase. Hence,
\be
D_{0,n}/\lambda_F \approx 1.1,\; 2.1,\; 3.1,\dots ~ (n=3/2,\;5/2,\;7/2,\dots).
\ee

These theoretical predictions are in agreement with the experimental results, $\overline{D}_n = (n-\epsilon_0)\lambda_F$ where $\epsilon_0\approx 0.32$. It is worth noting that the experimental average distance $\overline{D}_n$ is slightly greater than the equilibrium distance $D_{0,n}$ since the oscillation is not a pure sine curve. The non linearized interacting force $\bm{F}_1$ is slightly stronger for distances below the equilibrium distance $D_{0,n}$ than upper $D_{0,n}$. This slight distortion of the oscillations as it can be observed in Fig.~\ref{fexp-1}(a), means that drops spend more time in the oscillation at greater distance than at shorter distance.

The equilibrium distances are well predicted by looking at the local minima of the surface field. We can now go further and investigate the symmetrical oscillations around these quantized set of distances. 

\subsubsection{An analytical approach}
In this section, we rationalize the main experimental results in the "promenade mode", through an analytical approach. In order to get tractable analytical results, we make the following simplifying assumption that the drops, even in the oscillating regimes, remain in the neighborhood of these equilibrium distances. We can thus obtain a simplified theoretical expression for $\bm{F}_W=\bm{F}_1+\bm{F}_2$ by a linear expansion. 

The linearization of the forces acting on the drops proceeds in two steps. First, in situations of low memory, the interaction force is a function of the distance between the position of the drop 2 at $t_i$ and the position of the drop 1 at its previous impact $t_{i-1}$. This distance is noted $\ell_i$ as sketched in Fig.~\ref{fnotation} and differs slightly from $D_i$ by the vectorial equality
 \begin{equation}
 \bm{\ell}_i=\bm{D}_{i}-\bm{V}_i\, T_F
 \end{equation}
At the first order $D_i\gg V_i\, T_F$ (since the order of magnitude of $D_i$ is $\lambda_F$ and $V_i\, T_F$ is around $\lambda_F/50$) the equilibrium distances $\ell_{0,n}$ are given by $D_{0,n}$. The linearization of $\bm{F}_1$ around $D_{0,n}$ can be expressed as 
\begin{equation}
\bm{F}_1=- K_{\mathrm{int},n} (\ell_{i}-D_{0,n})\,\frac{\bm{\ell}_i}{\ell_i}.
\end{equation}
where $K_{\mathrm{int},n}$ is a spring-like constant which depends on the equilibrium distance $D_{0,n}$. A similar expression was used by Eddi \textit{et al.} \cite{zeeman}. Second, in situations of low memory, the self-interaction force $\bm{F}_2$ depends only on the distance between the past impact at $t_{i-1}$ and $t_i$. Consequently, $\bm{F}_2$ is only a function of the drop speed $\bm{V}_i$. The combined effect of the self-interaction force and the friction $\bm{F}_v$ can be expanded as 
\begin{equation}
\bm{F}_2+\bm{F}_v=- K_0(V_i-V_0)\, \dfrac{\bm{V}_i}{V_i}\,,
\end{equation}
where $K_0>0$. This means that $V_0$ is a stable point, and also the velocity of the free droplet. Moreover this expression is the lowest order expansion and is an asymptotic case of higher order theoretical expression \cite{StephMatt,oza}. Within the previous simplifying assumptions the dynamics of the system satisfies: 
\be \label{edyn2}
\dfrac{m_d}{T_F}\left( \bm{V}_{i+1} -\bm{V}_{i}\right)  = - K_0 (V_i-V_0)\, \dfrac{\bm{V}_i}{V_i} - K_{\mathrm{int},n} (\ell_{i}-D_{0,n})\,\frac{\bm{\ell}_i}{\ell_i}\,.
\ee
In the low memory regime, this equation is very general for a droplet influenced only by another one. The coefficients in this equation ($K_0$, $K_{\mathrm{int},n}$, $V_0$, and $D_{0,n}$) have values that can be determined from the dynamical coupling of a single drop with the standing wave it generates (\cite{JFM11,molacek2,oza}). $K_{\mathrm{int},n}$ is half of the the second order derivative of $f(D)$ around the equilibrium distance $D_{0,n}$; $K_0$ accounts for the linear stability of a self-propelled walker around its free velocity $V_0$ (see, {\it e.g.}, Ref. \cite{StephMatt} for their computation). In this paper they come from Eq.~(\ref{edyn}) and a Taylor expansion of $f(D)$ around $D=D_{0,n}$ and $D=0$, in which the free parameter $A_{\mathrm{eq}}$ allows a given free velocity $V_0$. In our computations, $V_0\approx 0.019\,\lambda_F/T_F$, $K_0\approx 0.11\, m_d/T_F$, and $K_{\mathrm{int},n} \approx 0.027\, m_d/T_F$ for instance for the equilibrium distance $D_{0,n}\approx 1.6\,\lambda_F$.\\

Although Eq. (\ref{edyn2}) can be numerically solved, we prefer to give insight into the promenade modes by providing an analytical resolution of the phenomena. In order to solve this equation analytically, several approximations are done as detailed in the Appendix~\ref{sappendix-B}. The main outputs concern both the longitudinal average velocity $\overline{V}_n$ of the bound pair as well as the characterization of the transverse oscillations. We first assume that $\overline{V}_n$ is constant and known, and within this assumption we can write the dynamics of the transverse motion around its equilibrium value $D_{0,n}$ (more precisely $Y_i = D_i - D_{0,n}$), as 
\bea \label{eYi}
\frac{Y_{i+2}-2Y_{i+1}+Y_i}{T_F^2} & -a\,\left[1-\frac{\left(Y_{i+1}-Y_i\right)^2}{\dot{Y}^2_c}\right]\frac{Y_{i+1}-Y_i}{T_F}\\
& +\, \omega^2\,Y_i\,=0\,,
\eea
where $a=(K_0/2\,m_d)\cdot\left[(1-\overline{V}^2_n/V^2_0)-2K_{\mathrm{int},n}\,T_F/K_0\right]$,  $\dot{Y}^2_c=8\,a\,m_d\,V^2_0/K_0$, and $\omega^2=2\,K_{\mathrm{int},n}/m_d$. Note that the coefficient 2 for $\omega^2$ comes from the reduced mass of the two-body problem.

This discrete-time dynamics appears to be similar to that of a Van der Pol like equation for the velocity. It is worth noting that the continuous limit of the discrete equation can be performed since the characteristic time of the dynamics, the oscillation period $T_{osc}$, is greater than around 20 times the inter-event time $T_F$ [see Fig.~\ref{fexp-2}(b)]. This implies, provided $a \ll 1$, a quasi-harmonic behavior of $Y_i$ at pulsation $\omega$. This dynamics can now be solved using continuous time. 

As discussed in the Appendix~\ref{sappendix-B}, since $\omega^2\,T_F>a$, the first term has to be expanded to the third order. Equation~(\ref{eYi}) can the be written in the more tractable form of a Van der Pol equation for the velocity:
\be \label{eYt}
\ddot{Y}-a'\,\left(1-\frac{\dot{Y}^2}{\dot{Y'}^{2}_c}\right)\,\dot{Y}+\omega^2\,Y(t)\,=0\,,
\ee
where $a'$ and $\dot{Y'}_c$ take into account the third order expansion of the first term in Eq.~(\ref{eYi}). It is important to note that the direct transposition of the discrete time dynamics Eq.~(\ref{eYi}) into a continuous time dynamics would have provided the same equation~(\ref{eYt}), but with the wrong coefficients. This technical point is detailed in the Appendix \ref{sappendix-B}. For this reason, it would have been wrong to go to the continuous limit without caution. The solution of Eq.~(\ref{eYt}) allows us to determine $\overline{V}_n$.\\

In order to test the validity of the model and the linearized approximation, we can compare the analytical results with the experimental measurements reported in Sec.~\ref{sexperiment}. In Fig.~\ref{fexp-1}(b) are given the theoretically predicted binding distances, in Fig.~\ref{fexp-2}(a) the average velocity of the center of mass along the $x$ axis, and in Fig.~\ref{fexp-2}(b) the period of the oscillatory motion. The theoretical binding distances are in excellent agreement with the experimental data. They are close to the minima of the Bessel function with a correction due to the exponential spatial extra damping. The amplitude of the analytical oscillations are weaker than the experimental ones owing to the linear limits in which the computation is done. This same reason is responsible for the shift to lower values of the oscillation period. This statement is confirmed by numerical simulations of the droplet's motion (in the same manner as Refs. \cite{JFM11,path-memory-pnas}) of the unlinearized problem: the results of the simulations are in better agreement with the experiments than the analytical ones.\\

In the following section we will consider the system from an energy viewpoint. As this is not easily feasible analytically we turned back to numerical simulations. In this framework the analytical results are recovered in a less simplified model. In particular a weak but non-zero memory can be taken into account.

\section{An energy equivalence \label{senergy}}

We can now turn to the initial question: is there an equivalence of viewpoint or at least a link between the kinetic energy of the coupled drops and the associated field ? Again, we consider only the horizontal motion of droplets. In this section, we first define the different energy terms involved in the coupling, from the drops' and from the waves' points of view. Then we compare the steady and oscillating kinetic energy to their wave counterpart. In this part, in order to compute the wave field, we will use numerical simulations using the principle established previously with values of the parameters established in Refs. (\cite{JFM11,molacek2,oza}). The memory remains short, $\Me = 4$.

\subsection{Interaction energy between two drops}
In the steady regime, two interacting droplets move in parallel along the $x$ axis at a velocity, $\overline{V}_n$, smaller than $V_0$ the free velocity of drops without any interaction. More precisely, for a long observation time the average velocity $\overline{V}_n$ depends on the average distance $\overline{D}_n$ between drops and converges towards $V_0$ when this distance (or the mode number $n$) increases, as shown in Fig.~\ref{fexp-2}(a). Thus, this loss of kinetic energy may suggest the existence of an effective binding energy. This energy is defined as the difference between the steady kinematic energies of the two drops with interaction, $\overline{E}_{n}$, and without interaction $E_0$:
\be \label{eEnergy}
\overline{E}_{n} = E_0 + E_{\mathrm{int},n}\,,
\ee

\be  \label{eEint}
  E_{\mathrm{int},n}= m_d\; (\overline{V}_{n}^2-V_0^2)\,.
\ee
In spite of the dissipative nature of the system, this equation permits the definition of an effective interacting energy.\\

The interaction between drops does not only change the linear velocity of drops $\overline{V}_{n}$, but also the the surface wave field. The typical wave field of an isolated walker exhibits at high memory an intrinsic interference pattern previously investigated in Refs.~\cite{JFM11,molacek2,oza} and shown in Fig.~\ref{fh-exp}(a). In the promenade modes, the superposition of the two wave patterns generates additional interference effects as observed in the experimental surface field in Figs.~\ref{fh-exp}(b)-\ref{fh-exp}(d) and for numerical simulations in Fig.~\ref{fh}. These interferences strongly depend on the binding distances $\overline{D}_n$ of the successive modes.

The surface density of a standing wave oscillating with the amplitude $h$ is proportional to $h^2$. Thus, we define the dimensionless energy of the standing waves at time $t_i$, by: 
\be \label{eH}
\displaystyle H(t_i) = \iint \left[h(\bm{r},t_i)\right]^2 \; d^2 \bm{r}\,,
\ee
where the integral is taken over the whole bath surface. In order to have a common energy definition for droplets that bounce in phase or in opposition of phase, the energy is averaged over the time $T_F$ of an oscillation, \text{i.e.}, $\left[h(\bm{r},t_i)\right]^2=1/2\,[h(\bm{r},t_i-T_F/2)]^2+1/2\,[h(\bm{r},t_i)^2]$, where $t=t_i$ is a time where a droplet (say, droplet $1$), bounces.

Let us write the field energy for two interacting and similar droplets, labeled 1 and 2. For the sake of simplicity, we first consider the low memory limit, for which we take into account only the last bounce [Eq \ref{equationbassememoire}]. The superposition of stationary circular waves is a very general feature of this system (with and without memory), and not only valid for one droplet [as in Eq.~(\ref{eh})]. So the relative surface height writes as $h(\bm{r},t_i)=h_1(\bm{r},t_i)+h_2(\bm{r},t_i)$, where $h_1$ and $h_2$, respectively, denote the height generated by the droplet $1$ and $2$, respectively. This permits writing the total surface energy $H(t_i)$ as a sum of a term issued from two drops without interaction, $H_{1}(t_i)+H_{2}(t_i)$
\be
\displaystyle H_{1}(t_i)+ H_{2}(t_i) = \iint \left[h_1(\bm{r},t_i)\right]^2 + \left[h_2(\bm{r},t_i)\right]^2 \; d^2 \bm{r} 
\label{eH0}
\ee
and an interaction energy
\begin{equation}
\displaystyle H_{1-2}(t_i)=2 \iint  h_1(\bm{r},t_i)h_2(\bm{r},t_i) \; d^2 \bm{r} \,. 
\label{eHint}
\end{equation} 
Note that the interacting term is an interference term resulting from the overlap of two fields, $h_1(\bm{r},t_i)$ and $h_2(\bm{r},t_i)$.

The interaction between droplets changes trajectories of free droplets. Nevertheless, when we take into account only the last bounce of each droplet, $H_{1}(t_i)+ H_{2}(t_i)$ is equal to $H_0$, the wave energy of two free droplets. Thus, in the low memory limit, $H_{1-2}(t_i)$ encapsulates the whole wave interaction energy, $H_{\mathrm{int}}$, between droplets. When the memory is taken into account, the modified trajectories imply that $H_{1}(t_i)+ H_{2}(t_i)\neq H_0$. Hence we again define the wave interaction energy from the steady state as the difference between the wave energy, $H$, of the steady trajectories and the wave energy of the two droplets without any interaction, $H_0$:
\be 
H_{\mathrm{int}} = H - H_0\,.
\ee
For long observation time, the trajectories of the two droplets in a promenade mode are characterized by two parameters, the distance $D$ between them and their velocity $\overline{V}$. However, at low memory ($\Me=4$), numerical calculations show that $H_{\mathrm{int}}$ depends strongly on the distance between droplets and weakly on their velocity (when the latter is in the range of the free velocity $V_0$) as it is observed in Fig.~\ref{fH}(a). This enables writing the wave interacting energy as a function of only one parameter, $H_{\mathrm{int}}=H_{\mathrm{int}}(D)$.

It is now interesting to look for a possible relation between the two different definitions of the interaction energy: the one from the particle point of view [see Eq.~(\ref{eEint})] and the other from the wave point of view [see Eq.~(\ref{eHint})]. Since the dynamics consists of a steady motion along the $x$ axis, associated with transverse and longitudinal oscillations, we compare the energies of the two different points of view at long and short timescale. Let us first show that the wave interacting energy allows us to retrieve the quantization of the promenade modes.

\subsection{Equilibrium distance}
\begin{figure}
\centering
\includegraphics[width=1 \columnwidth, clip=true]{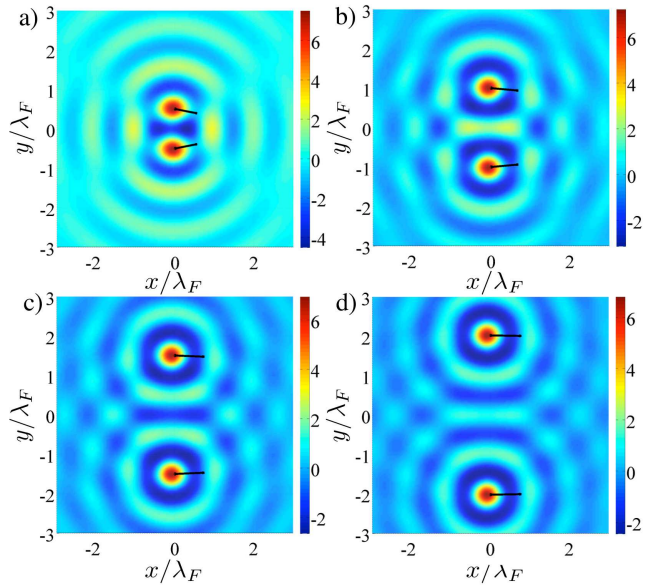}
\caption{\small (Color online) Four wave fields computed for (a) $\overline{D}_1=0.6$, (b) $\overline{D}_2=1.6$, (c) $\overline{D}_3=2.6$ and (d) $\overline{D}_4=3.6$. Positions where the drops collide with the surface bath and the velocity of droplets are indicated by crosses and vectors. The length of the vector is equal to 50 times the distance traveled by the walker between two successive impacts. Colorbars are expressed in $10^{-3}\lambda_F$. \label{fh}}
\end{figure}

\begin{figure}
\centering
\includegraphics[width=0.85 \columnwidth, clip=true]{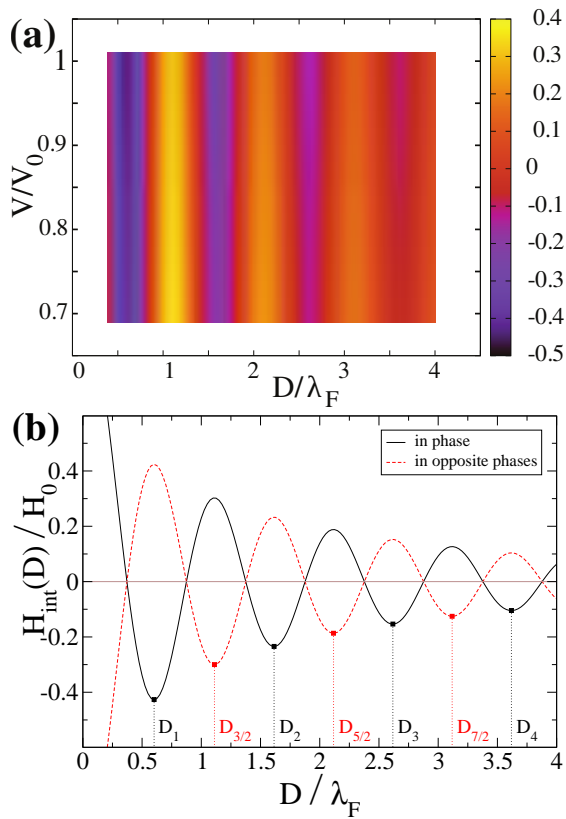}\\
\includegraphics[width=0.85 \columnwidth, clip=true]{Fig6b.eps}
\caption{\small (Color online) The energy $H_{\mathrm{int}}$ stored in the wave field for the steady state. (a) Heat map of $H_{\mathrm{int}}/H_0$ for two droplets bouncing in phase as a function of the steady velocity $\overline{V}$ of droplets and the distance $D$ between them. (b) $H_{\mathrm{int}}/H_0$ as a function of one parameter, the distance $D$ separating the two drops which bounce in phase or in opposites phases, when $\overline{V}=V_0$. The relative minima $H_{\mathrm{int}}(D_n)/H_0$ at distances $D_n$ are shown.}
\label{fH}
\end{figure}

The existence of a coupled motion has been theoretically investigated in the previous section. The quantized distance corresponds to the position at which the mutual interaction force is zero. It can be analyzed from an energy point of view averaged over a long time duration. Figure~\ref{fH}(b) shows the evolution of the wave interaction energy, $H_{\mathrm{int}}(D)$, as a function of the distance $D$ between the two drops. Its minima, $D_n$, are very close to equilibrium distances $D_{0,n}$ reported in the previous section. Note that the drops can be in phase or in opposite phases which shifts the position on the minima. The related wave field is plotted in Fig.~\ref{fh}(b) and shows qualitatively that these distances correspond to a destructive interference between the waves generated by the two coupled drops.

In order to present the main physical features, the interaction can be analytically computed in an asymptotic limit of low memory (labeled ``$\mathrm{asymp}$") as given by Eq~(\ref{equationbassememoire}). Note that only the distance $D$ between droplets is relevant for the wave energy in this limit, in which only the last bounce is taken into account. In this limit, the wave interacting energy writes (see the Appendix~\ref{sappendix-C}) as
\begin{equation}
\frac{H_{\mathrm{int}}^{\mathrm{asymp}}}{H_0}=\mathrm{J_0}\left(2\pi \dfrac{D}{\lambda_F}\right).
\label{Eintgraf}
\end{equation}
Here, in Eq. (\ref{Eintgraf}), we have neglected the spatial viscous damping of standing waves and the finite velocity propagation of signals. Equation (\ref{Eintgraf}) mainly relies on the Graf's decomposition theorem \cite{fcspeciales}.

The quantization of the drop distance is similar to that which would have been obtained from the minimization of the surface energy. We can now finally turn to the question initially raised: What is the link between these surface energies, --a wave point of view,-- and the kinetic energy of the two drops, --a particle point of view?  

\subsection{Energy equivalence}
\begin{figure}
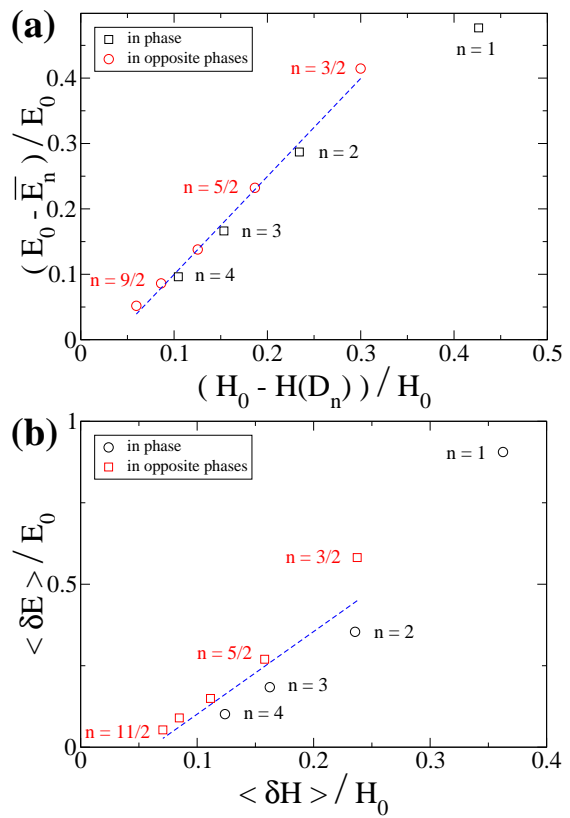

\centering
\includegraphics[width=0.85 \columnwidth, clip=true]{Fig7a.eps}\\
\includegraphics[width=0.85 \columnwidth, clip=true]{Fig7b.eps}
\caption{\small (Color online) The relation of the non-dimensional kinetic energy of the droplets with the energy stored in the wave field as obtained from the numerical simulation. (a) Steady regime. (b) Oscillatory motion. The dashed lines show a linear fit. \label{equivalence}}
\end{figure}
The steady speed of the coupled drops $\overline{V}_n$ is significantly lower than their free speed $V_0$, \textit{i.e.}, their speed in the absence of mutual interaction. They interact through their waves and the resulting state can be described, at least in principle, either by the kinetic energy for the coupled system or by the amount of energy stored in the surface. We plot in Fig.~\ref{equivalence}(a) the relative loss of kinetic energy with respect to the relative loss of wave interaction energy in the corresponding situation, {\it i.e.}, $E_{\mathrm{int},n}/E_0$ as a function of $H_{\mathrm{int}}(D_n)/H_0$ for different modes $n$. 

The plot of the kinetic energy of the two drops as a function of the energy of the wave field shows that they are related by an affine and not a linear relation. In order to understand this effect one must remember that the bound states in the promenade modes are locally stable but globally unstable. When a promenade mode of order $n$ is disturbed, the two walkers do not naturally fall toward a lower order of a promenade mode but they tend to increase their average distance in another bound state, or else they tend to become two free walkers. It is worth noting that this phenomenon is different to usual conservative systems, for which the disturbed systems tend to evolve toward lower energy states. In our case, the disturbed bound states evolve toward higher energy states, both for the kinetic energy of droplets and for the wave energy. Moreover, specifically due to the spatial damping of standing waves, for which the characteristic length is $\delta=2.5 \lambda_F$ [see Eq.~(\ref{eh})], the field generated by a walker becomes very small at distances greater than $\delta$. This strengthens the non-stability of possible bound pairs of walkers when the binding distance between them increases. Lastly, this locally stable but globally unstable phenomenon can be seen as an additional positive energy (stored in the wave field) that a bound pair of walkers has to overcome in order to become unbound, as can be seen in Fig.~\ref{equivalence}(a).

The proportion relation between the interaction energy defined by means of the kinetic energy of droplets and the wave interaction energy is the main, and {\it a priori} surprising, result. Furthermore the coefficient of proportionality is of the order of one. In our numerical computation, the wave energy is of the order of $10^{-9}\:\mathrm{J}$, greater than about 10 times the kinetic energy of droplets, $E_0=m_d\,V_0^2$. The wave energy (with the previously-omitted prefactor) is computed as for monochromatic plane waves $H_0 = 1/2\left[\rho\,g+\sigma\,(2\pi/\lambda_F)^2\right]\iint h^2(\bm{r}) \, d^2 \bm{r}$, where $\rho$ is the volumetric mass density of the liquid bath and $\sigma$ its surface tension \cite{FluidMech}. In order to strengthen this result, we note that very similar plots are obtained from other definitions of the minimal wave interaction energy: (1) when the wave energy is evaluated from droplets moving parallel with a distance and a velocity equal to the corresponding averaged quantities over promenades trajectories of droplets [{\it i.e.}, $\overline{D}_n$ and $\overline{V}_{n}$ as in Figs~\ref{fexp-1}(b) and \ref{fexp-2}(a)], or (2) $H_{\mathrm{min}}$, the minimal value of the wave energy $H(t)$ along the corresponding promenade trajectory of droplets.

We now turn the investigation toward shorter observation times; {\it i.e.}, we now consider oscillatory motions, transverse and longitudinal ones. Fig.~\ref{equivalence}(b) plots the relative gain of fluctuating kinetic energy with respect to the relative gain of the wave energy around the minimum. More precisely, we plot $\langle \delta E \rangle/E_0$ (where $\delta E = m_d\, [ V_y^2+(V_x-\overline{V}_n)^2]$) as a function of the corresponding situation of $\langle \delta H(t) \rangle/H_0$ (where $\delta H(t) = H(t) - H_{\mathrm{min}}$, in which $H_{\mathrm{min}}$ denotes the minimal value of the wave energy $H(t)$ along the corresponding promenade trajectory of droplets). The time averaging operation $\langle . \rangle$ is realized over one period of oscillation. The relation between these two different fluctuating energies is similar to that observed for the interaction energies. Moreover the coefficient of proportionality is again equal to one in order of magnitude.

\section{conclusion}

We have studied here a dynamical association of a droplet with a physical wave field. In the resulting ``walker,'' the two components have a non-dissociable link with each other. The two droplets influence one another by means of the standing waves they both generate. In spite of being a dissipative structure, a single isolated walker has a steady regime of motion. The process by which it receives energy from the forcing is complex. Both the drop and the waves receive energy directly from the forcing: the droplet is kicked upward at each collision with the interface, and the wave is partly sustained by parametric forcing. There are also energy exchanges between the two components. When it falls onto the fluid interface the drop transfers some of its energy to the bath where it generates a new wave. As its lifts, the drop receives a horizontal motion by a transfer from the wave. Linked with these energy exchanges there is also information interplay: the drop generates the wave-field and this wave-field determines the direction of the drops. 

The global energy of the system had not hitherto been explored. For this purpose we have used here the existence of bound walkers specifically focusing on the promenade modes that are the simplest binding modes in which two walkers move parallel to each other. The experiments show that when the distance between two identical walkers is reduced, they stabilize in one of several steady regimes characterized by a parallel mean propagation in which case their mean distance is quantized. Their translation velocity is found to increase with the binding distance. We have computed the evolution with the binding strength of both the kinetic energy of the drops and the energy content of the whole wave-field. The computation of the two energies shows that they are of the same order of magnitude. More importantly they are found to be proportional to each other. These linear relations do not correspond to a transfer between a kinetic and surface energy term which would have involved a linear relation with a negative sign. Here this equivalence is the signature, in the energetic domain, of the dual nature of the walker that can be described either by its path or by its corresponding surface wave .

\begin{acknowledgments}
We thank St\'ephane Perrard for discussion and comments. This research is supported by the French Agence Nationale de la Recherche through the project ``ANR Freeflow'', and the Labex WIFI (ANR-10-LABX-24) within the French Program Investments for the Future under reference ANR-10-IDEX-0001-02 PSL and the AXA Research Fund. 
\end{acknowledgments}

\appendix 
\section*{Appendix}
\renewcommand{\theequation}{A\arabic{equation}}
\setcounter{equation}{0}  
\renewcommand{\thesubsection}{A}
\subsection{The dynamics of a droplet as resulting from effective forces \label{sappendix-A}}
According to the theoretical framework used in Section~\ref{stheoretical}, there are two kinds of forces which govern the evolution of bouncing positions of a droplet: (1) the incremental velocity `kick' that the drop receives when it collides with the bath surface, and (2) a viscous-like force that the liquid bath induces on the drop. It can be noted that the horizontal velocity of the droplet is assumed to be continuous when it leaves the bath surface and starts its free flight. 

\begin{figure}
\centering
\includegraphics[width=1 \columnwidth, clip=true]{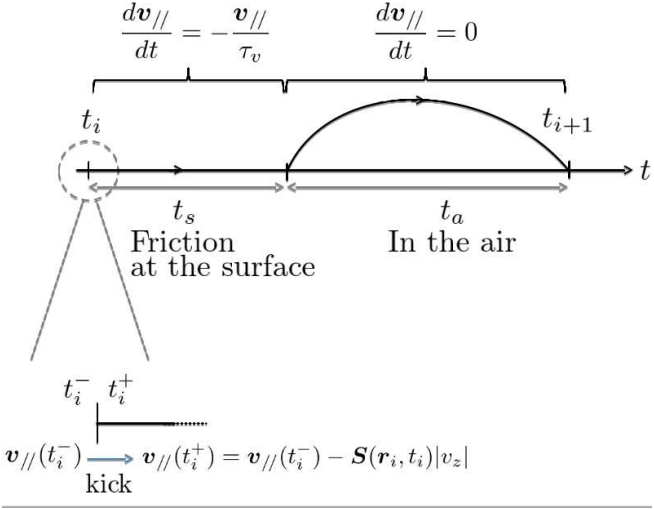}
\caption{\small Notations and the dynamics of the bouncing droplet used in the computation.}
\label{fnotation-append}
\end{figure}

Let respectively $t_s$ and $t_a$ (see Fig.~\ref{fnotation-append}), be the fraction of time spent by the droplet in contact with the bath and in free flight ($t_s+t_a=T_F$). The droplet interacts on the liquid bath with an apparent friction time $\tau_v$ (see Refs. \cite{path-memory-pnas,oza,molacek2}) which accounts for the loss of the kinetic energy of the drop at the surface, while the droplet is considered to have a free motion in the air. It can be noted that the friction time, $\tau_v$, depends on the mass of the droplet~\cite{molacek2,oza}. Let $\vpara(t_i^-)$ (respectively $\vpara(t_i^+)$) being the instantaneous velocity parallel to the horizontal plane of the drop, just before (respectively just after) the impact between the drop and the bath, at the time $t_i$. Thus
\be \label{evpara-append}
\vpara(t_{i+1}^-) = \vpara(t_i^+ +t_s) = \exp(-\frac{t_s}{\tau_v}) \cdot \vpara(t_i^+)\,.
\ee
Moreover, when the droplet hits the bath surface, it receives an incremental velocity `kick' which can be considered as the result of a soft shock of a drop on a slippery surface bath. The incremental velocity kick, at position $\vectr_i$ and at time $t_i$ where the drop collides with the liquid bath, is then oppositely proportional to the slope $\bm{S}(\vectr_i,t_i) = \left[\bm{\nabla}_r h\right](\vectr=\vectr_i,t_i)$ of the relative surface height $h(\vectr_i,t_i)$.  The study takes place in the weak slope limit (which is experimentally confirmed). This yields, in the limit of small slope, to 
\be \label{ekick-append}
\vpara(t_i^+) = \vpara(t_i^-) - \bm{S}(\vectr_i,t_i)\cdot |v_{z}|\,,
\ee
where $v_{z}$ denotes at time $t_i$ the relative velocity in the vertical axis between the fall of the drop and the oscillating bath.

Let $\bm{V}_i=(\vectr_{i}-\vectr_{i-1})/T_F$ be the velocity vector between two bounces of a droplet from the time $t_{i-1}$ to $t_i$. $\bm{V}_{i}$ is collinear to $\vpara(t_i^-)$, with the proportionality coefficient being $c=t_a+\tau_v(e^{t_s/\tau_v}-1)$. According to Eqs.~(\ref{evpara-append}) and (\ref{ekick-append}), we can write the evolution of velocity vector as resulting from effective forces, such that:
\be \label{edyn-append} 
\dfrac{m_d}{T_F}\left(\bm{V}_{i+1} - \bm{V}_{i}\right) = \vectfv + \vectF_W \,,
\ee
where
\be \label{efv-append}
\vectfv=-\dfrac{m_d}{T_F}C_v \cdot \bm{V}(t_i),~~\mathrm{with}~~C_v=1-e^{-t_s/\tau_v}>0\,,
\ee
and 
\be \label{eF-append}
\vectF_W = - \dfrac{m_d}{T_F}C_F \cdot \bm{S}(\bm{r}_i,t_i),~~\mathrm{with}~~C_F=c\, |v_{z}|\, e^{-t_s/\tau_v}>0\,.
\ee

\renewcommand{\thesubsection}{B}
\subsection {Solving the dynamics of the promenade mode \label{sappendix-B}}
This appendix aims at resolving analytically Eq.~(\ref{edyn2}), in addition to the evolution of the transverse distance $D_i$ between droplets, $D_{i+1} = D_{i} + 2\, V_{y,i+1}T_F$, provided that some assumptions are given. 

First, we assume that $\ell_{x,i}=V_{x,i}T_F$ is negligible compared with $\ell_{y,i}$. We recall that experimentally, $D_i$ is of the order of magnitude of the Faraday wavelength $\lambda_F$, and $V_x$ and the free velocity of the drop, $V_0$ are of the order of magnitude of $\lambda_F/(50\, T_F)$), and $V_{i}=V_0\,\sqrt{1+\left(\frac{V^2_{y,i}+V^2_{x,i}-V^2_0}{V^2_0}\right)}$ can be expanded at first order in the term in brackets. 

Second, Eq.~(\ref{edyn2}) projecting onto the $x$ and $y$ axes provides two coupled differential equations for the speed in the $x$ and $y$ directions. Nevertheless, in this problem, it appears from numerical calculations that $V_x$ evolves more smoothly than $V_y$, which is also confirmed experimentally. Thus we assume that $V_{x,i}\approx \overline{V}_n$, {\it i.e.}, $V_x$ weakly oscillates around its average value $\overline{V}_n$. This leads to writing the dynamics of the transverse motion around its average value, $Y_i=D_i-D_{0,n}$, as
\bea \label{eYi-append}
\frac{Y_{i+2}-2Y_{i+1}+Y_i}{T_F^2} & -a\,\left[1-\frac{\left(Y_{i+1}-Y_i\right)^2}{\dot{Y}^2_c}\right]\frac{Y_{i+1}-Y_i}{T_F}\\
& +\, \omega^2\,Y_i\,=0\,,
\eea
where $a=(K_0/2\,m_d)\cdot\left[(1-\overline{V}^2_n/V^2_0)-2K_{\mathrm{int},n}\,T_F/K_0\right]$, $\dot{Y}^2_c=8\,a\,\,m_d\,V^2_0/K_0$ and $\omega^2=2\,K_{\mathrm{int},n}/m_d$. 

This discrete-time dynamics appears to be similar to that of a Van der Pol like equation, the velocity being controlled instead of the amplitude. This implies, because $a \ll 1$, a quasi-harmonic behavior of $Y_i$ with the pulsation $\omega$. This dynamics can now be solved using continuous time. Nevertheless, since $\omega^2\,T_F>a$ the first term should be developed to the third order. [Since $a\ll 1$ we expect that $Y(t)\propto \cos(\omega\,t\,+\phi_0)$, thus $|T_F\,\dddot{Y}(t)|>|a\,\dot{Y}(t)|$ when $\omega^2\,T_F>a$.] Thus:
\be \label{eYt3-append}
T_F\,\dddot{Y}+\ddot{Y}-a\,\left(1-\frac{\dot{Y}^2}{\dot{Y}^2_c}\right)\,\dot{Y}+\omega^2\,Y(t)\,=0\,.
\ee
According to what precedes, the latter equation is reduced by assuming that $\dddot{Y}(t) \approx -\omega^2\,\dot{Y}(t)$, giving a more tractable equation:
\be \label{eYt-append}
\ddot{Y}-a'\,\left(1-\frac{\dot{Y}^2}{\dot{Y'}^{2}_c}\right)\,\dot{Y}+\omega^2\,Y(t)\,=0\,,
\ee
where $a'=a+\omega^2\,T_F$ and $\dot{Y'}^{2}_c=8\,a'\,m_d\,V_0^2/a\,K_0$. Differentiating Eq.~(\ref{eYt-append}) and writing $Z=\dot{Y}$ and $Z^2_c=\dot{Y'}^{2}_c/3$,  it yields the usual Van der Pol equation, $\ddot{Z}-a'\,\left(1-\frac{Z^2}{Z^2_c}\right)\,\dot{Z}+\omega^2\,Z(t)=0$, whose solution is, since $a' \ll 1$ (see, e.g., Ref. \cite{ordre-dans-chaos}), $Z(t)=2\sqrt{Z^2_c}\,\sin(\omega\,t+\phi_0)$. So, according to the previous assumptions, the distance $D$ between two droplets harmonically oscillates around the average distance $D_0$, with the pulsation $\omega$ and the amplitude $A_Y=\frac{2\sqrt{\dot{Y'}^{2}_c}}{\omega\sqrt{3}}$. 

Finally, knowing $Y(t)$, and thus $V_{y}(t)$, the projection of Eq.~(\ref{edyn2}) onto the x-axis and solving $\langle (V_{i}-V_0)/V_{i}\rangle = 0$ provides the the average velocity $\overline{V}_n$. Indeed, according to what precedes, $\langle \frac{V_{i}-V_0}{V_{i}}\rangle = 0$ yields $\frac{2}{\pi}\frac{V_0}{\sqrt{\overline{V}_n^2+\langle V^2_y\rangle}}\cdot K(k)=1$, where $\langle V^2_y\rangle=\frac{\dot{Y'}^{2}_c}{6}$ and $K(k)$ means the complete elliptic integral of the first kind~\cite{handbook-math} with $k^2=\frac{\langle V^2_y\rangle}{\overline{V}_n^2+\langle V^2_y\rangle}$.

\renewcommand{\thesubsection}{C}
\subsection {Asymptotic limit of the wave interaction energy \label{sappendix-C}}
In this appendix we analytically compute the wave interaction energy in the limit of low memory and no viscosity of the liquid bath. Let two droplets, labeled 1 and 2, being respectively at position $\bm{r}^{(1)}$ and $\bm{r}^{(2)}$ at time $t_i$, at the distance $D=\Vert \bm{r}^{(1)}-\bm{r}^{(2)}\Vert$ between them. The relative height emitted by the droplet $j$ ($j=1$ or 2) is given by Eq.~(\ref{equationbassememoire}), where $\delta\to + \infty$, here written as $h_j(\vectr) = A_{\mathrm{eq}} \mathrm{J}_0\left[2\pi\frac{\Vert\vectr-\vectr^{(j)}\Vert}{\lambda_F}\right]$; and from Eq.~(\ref{eH0}) the wave energy of the two droplets without interaction becomes
\begin{equation} \label{eH0-append}
\displaystyle H_{0}^{\mathrm{asymp}} = 2\, A_{\mathrm{eq}}^2 \iint_\mathcal{S} \left[\mathrm{J}_0\left(2\pi\frac{\Vert\vectr-\vectr^{(1)}\Vert}{\lambda_F}\right)\right]^2 \; d^2 \bm{r}\,,
\end{equation}
where $\mathcal{S}$ means a disk of radius $R \gg \lambda_F$ centered in $(\bm{r}^{(1)} + \bm{r}^{(2)})/2$. Note that in the main text where the damping of progressive capillary-gravity waves is taken into account, the integral is taken over the whole surface, and in practice over a disk of radius around seven times the characteristic damping distance $\delta$.

According to Eq.~(\ref{eHint}) the wave energy interaction writes in this case as 
\be
\displaystyle H_{\mathrm{int}}^{\mathrm{asymp}}= 2\, A_{\mathrm{eq}}^2 \iint_\mathcal{S} \mathrm{J}_0(2\pi\frac{\Vert\vectr-\vectr^{(1)}\Vert}{\lambda_F})\, \mathrm{J}_0(2\pi\frac{\Vert\vectr-\vectr^{(2)}\Vert}{\lambda_F}) \, d^2 \bm{r}\,.
\label{A10}
\ee
Using the Graf’s decomposition theorem \cite{fcspeciales} yields,
\bea
\displaystyle & \mathrm{J}_0(2\pi\frac{\Vert\vectr-\vectr^{(2)}\Vert}{\lambda_F})=  \mathrm{J}_0(2\pi\frac{\Vert\vectr-\vectr^{(1)}\Vert}{\lambda_F})\: \mathrm{J}_0(2\pi \frac{D}{\lambda_F})\\
& + 2 \sum_{n=1}^{+ \infty} \mathrm{J}_n(2\pi{\Vert\vectr-\vectr^{(1)}\Vert}{\lambda_F})\: \mathrm{J}_n(2\pi \frac{D}{\lambda_F}) \cos(n\,\theta)
\eea
where $\mathrm{J}_n$ indicates the Bessel function of the first kind of order $n$ and $\theta$ is the angle between $\bm{r}^{(2)}-\bm{r}^{(1)}$ and $\bm{r}-\bm{r}^{(1)}$. Once inserted in Eq.~(\ref{A10}), every terms $n \geq 1$ get erased by the polar integration, the wave energy interaction becomes
\bea
\displaystyle H_{\mathrm{int}}^{\mathrm{asymp}}= 2\, A_{\mathrm{eq}}^2\: \mathrm{J}_0(2\pi\frac{D}{\lambda_F}) \iint_\mathcal{S} \left(\mathrm{J}_0(2\pi\frac{\Vert\vectr-\vectr^{(1)}\Vert}{\lambda_F})\right)^2 \; d^2 \bm{r}\,.
\eea
Hence, Eq.~(\ref{A10}) becomes
\begin{equation}
H_{\mathrm{int}}^{\mathrm{asymp}}/H_{0}=\mathrm{J_0}\left(2\pi \frac{D}{\lambda_F}\right).
\end{equation}


\end{document}